\documentclass[12pt]{article}
\usepackage{epsfig,amssymb,cite,multirow} 
\usepackage{amsmath}

\usepackage{amscd}
\usepackage{amsmath,graphicx}
\usepackage{xypic}
\usepackage[matrix,arrow,curve]{xy}
\usepackage{graphicx}
\usepackage{verbatim}
\usepackage{epsf}
\usepackage{latexsym}
\usepackage{amsmath,amsfonts,amssymb,amsthm}
\usepackage{amsmath,amsthm}

\def\bseq{\begin{subequation}}  % = 1a 1b
\def\eseq{\end{subequation}}
\def\bsea{\begin{subeqnarray}}  % = 1.1a 1.1b
\def\esea{\end{subeqnarray}}
\newcommand{\bbox}{\lower.2ex\hbox{$\Box$}}

\newcommand{\beq}{\begin{equation}}
\newcommand{\eeq}{\end{equation}}
\newcommand{\bea}{\begin{eqnarray}}
\newcommand{\eea}{\end{eqnarray}}
\newcommand{\ena}{\end{eqnarray}}

\renewcommand{\]}{\right]}

\newcommand{\be}{\begin{equation}}
\newcommand{\ee}{\end{equation}}

\begin{document}
\setcounter{page}{0}
\begin{titlepage}
\titlepage
\begin{flushright}
UCSD-PTH-10-05\\
LPTENS-10/41\\
\end{flushright}
\begin{center}
\LARGE{\Huge 
Additional Light Waves}\\
\Huge{in Hydrodynamics and Holography}
%Pekar's waves  in Holographic Optics}

\end{center}
\vskip 1.5cm \centerline{{\bf Antonio Amariti$^{a}$\footnote{\tt amariti@ucsd.edu}, Davide Forcella$^{b}$\footnote{\tt forcella@lpt.ens.fr}, Alberto Mariotti$^{c}$\footnote{\tt alberto.mariotti@vub.ac.be} 
%..................$^{c}$ \footnote{\tt ........................}
}}
%\medskip
\vskip 1cm
\footnotesize{

\begin{center}
$^a$Department of Physics, University of California,
San Diego\\
9500 Gilman Drive, La Jolla, CA 92093-0319, USA
\\
\medskip
$^b$ Laboratoire de Physique Th\'eorique de l'\'Ecole Normale Sup\'erieure \\
and CNRS UMR 8549\\
24 Rue Lhomond, Paris 75005, France
\\
\medskip
$^c$
Theoretische Natuurkunde, Vrije Universiteit Brussel \\
and
The International Solvay Institutes\\ 
Pleinlaan 2, B-1050 Brussels, Belgium\\
\end{center}}

\bigskip

\begin{abstract}
We study
the phenomenon of additional light waves (ALWs), observed in crystal optics:
two or more electromagnetic waves with the same
 polarization, but different refractive index, 
propagate simultaneously in a isotropic medium.
We show that ALWs are common in relativistic hydrodynamics,
and in particular in strongly coupled systems that admit a
dual gravitational description, where the ALWs are dual
to quasi normal modes in the AdS gravity.
We study both the transverse and the longitudinal light 
wave propagation. In the longitudinal channel 
we find a transition between regimes with
different number of excitonic resonances which resembles the transition 
to standard optics observed in  crystals. 
\end{abstract}

\vfill
\begin{flushleft}
{\today}\\
\end{flushleft}
\end{titlepage}

\newpage

\tableofcontents

%%%%%%%%%%%%%%%%%%%%%%%%%%%%%%%%%%%%%%%%%%%%%%%%%
%%%%%%%%%%%%%%%%%%%%%%%%%%%%%%%%%%%%%%%%%%%%%%%%%
%%%%%%%%%%%%%%%%%%%%%%%%%%%%%%%%%%%%%%%%%%%%%%%%%
%%%%%%%%%%%%%%%%%%%%%%%%%%%%%%%%%%%%%%%%%%%%%%%%%
%%%%%%%%%%%%%%%%%%%%%%%%%%%%%%%%%%%%%%%%%%%%%%%%%
%%%%%%%%%%%%%%%%%%%%%%%%%%%%%%%%%%%%%%%%%%%%%%%%%
%%%%%%%%%%%%%%%%%%%%%%%%%%%%%%%%%%%%%%%%%%%%%%%%%
%%%%%%%%%%%%%%%%%%%%%%%%%%%%%%%%%%%%%%%%%%%%%%%%%

\section*{Introduction}
\addcontentsline{toc}{section}{Introduction}

In the recent years an intense research on light  
propagation in some artificial media, called metamaterials,
has been developed \cite{Smith1}.
Some very uncommon properties, theoretically predicted, have become viable.

One of the most attractive  is
the negative refraction of light,
originally predicted in \cite{Veselago}. It is
the phenomenon for which the phase velocity and energy flux 
of a electromagnetic wave in a material have opposite directions of propagation.
One may ask if there exist any generic class of media,
for which the EM response functions are exactly computable, that have a negative refractive index.
With this motivation, in \cite{Amariti:2010jw}, we studied negative refraction 
in relativistic hydrodynamics and in particular in strongly coupled relativistic media
We established a new connection
between optics and geometry through the gauge/gravity correspondence 
\cite{Maldacena:1997re}, and we observed that negative refraction
is quite ubiquitous for  this class of media.
The negative refraction of light appears because the
dispersion relations are dissipative and they violate the locality, 
i.e. there is spatial dispersion.

Spatial dispersion is an interesting property of media 
that leads to 
new amazing non-standard optics.
In media with spatial dispersion 
there is a linear 
non local relation between
the electric field and the electric induction. This translates in
the dependence of the permittivity on
the wave vector.

A non standard optic phenomenon associated to spatial dispersion is
the propagation of additional
light waves (ALWs) in crystal optics. 
The existence of ALWs has been predicted in 1957 by   Pekar \cite{Pekar}.
He claimed that 
spatial dispersion effects are relevant in
presence of  resonances given by electron-hole bound states,
called excitons.
At the resonance frequency the
interaction among the excitons and the photons is strong, and the
light wave that propagates in the material is a mixture of these 
states, called polaritons. 
The energy necessary to create the resonance is proportional 
to the momentum of the incoming photon. 
As a consequence the linear relation between the 
electric field and the electric induction violates locality, and 
the permittivity has a pole that depends on the wave vector.

In absence of strong spatial dispersion the dispersion equation has 
a single solution for $n^2$, the square of the refractive index. 
If there are strong spatial dispersion effects new solutions
to the dispersion relation arise. 
At the resonance frequency these new solutions are associated to the propagation 
of ALWs. 
This exotic picture, in which many different propagating
light waves are associated to the same incoming photon, 
has been subsequently observed in
experiments (see \cite{Strashnikova} and references therein).

The lifetime of an exciton depends on the order parameters of the theory.
If its lifetime  becomes  too short, the ALW associated to this resonance
does not propagate anymore. 
Pekar's theory predicts that by varying the order parameters
the number of waves propagating in the medium varies too.
This transition among different optical regimes has been observed in
crystals with one exciton.
In that case for low temperature the lifetime of the exciton
is large enough for both waves to propagate. 
As the temperature increases the exciton lifetime reduces
and above a critical value only the standard single wave propagates.

In this paper we investigate the existence of ALWs in relativistic hydrodynamics 
and in particular in strongly coupled media with a gravity dual description. 
In such media the propagation of light, 
in the  hydrodynamical regime, typically shows spatial dispersion.
We revisit the model studied in \cite{Amariti:2010jw}, where the refraction
of transverse light waves was studied.
It was found that at finite temperature and
chemical potential the refractive index was 
negative in the low frequency and small
wave vector-regime.
The refractive index was computed by considering only the 
lowest order dependence of the permittivity from
the wave-vector. Here 
we observe that if we take into account the whole
spatial dispersion effects
there is a second solution of the dispersion equation, as 
in the case of crystals.
We study this  additional light wave and we show that
it propagates for every value of the order parameters.

Then we analyze the propagation of longitudinal electromagnetic
waves in the same system.
In absence of spatial dispersion longitudinal light waves do not propagate in 
a medium.  On the contrary here the permittivity is strongly dependent
on the wave-vector, and there is a propagating wave
at finite temperature and zero charge density. 
If a chemical potential is turned on, there is another solution to the dispersion
equation, and hence an ALW. 
The analysis of the propagation of these
two waves depends on the values of the order parameters. Indeed,
by varying them, there is an optical transition,
between the regimes with one and two propagating waves.
This is similar to the transition to standard optics predicted 
by Pekar's theory.

The paper is organized as follows.
In section \ref{sec1} we review the linear response theory of media with spatial dispersion.
In section \ref{sec2} we discuss the propagation of the light waves in media with 
spatial dispersion and we review the basic properties of Pekar's theory. 
In section \ref{sec3} we observe that if the medium  
is described by relativistic hydrodynamics, then ALWs are generated. 
In section \ref{sec4} we study the ALWs in an holographic theory 
in the hydrodynamical regime. In section \ref{sec5}
 we study some properties of these waves and then we conclude.
In the Appendix \ref{app1} we discuss the relation between the current correlators and
the linear response function of a medium with spatial dispersion. In the Appendix
\ref{app2} we  discuss  the Poynting vector and its relation with the sign of the refractive 
index for the ALWs.
\section{Linear response and spatial dispersion}
\label{sec1}

In this section we review the linear response electromagnetic functions for media with
spatial dispersion \cite{Agranovich,Agranovichlibro,Landau8,Landau10} i.e. we take into 
account the  momentum dependence. 
In this case it is common to use the $E,B,D$ approach to macroscopic electromagnetism for 
which the Maxwell equations in the Fourier space 
are
\be{}\label{Maxwell}
k \cdot B=0 \qquad,\qquad k \cdot D=0 \qquad,\qquad k \wedge E =  \omega B \qquad,\qquad k \wedge B =- \omega D 
\ee
and $D_i=\epsilon_{ij}(\omega,k) E_j$. 
The linear response
of the medium  is encoded in the permittivity tensor, 
$\epsilon_{ij}(\omega,k)$. For isotropic media this tensor naturally decomposes in transverse 
$\epsilon_T (\omega,k)$ and longitudinal part $\epsilon_L(\omega,k)$, and
\be{}
D_i=\epsilon_{ij}(\omega,k) E_j = \epsilon_T (\omega,k) E_i^T + \epsilon_L(\omega,k)  E_i^L
\ee
where $E_i^T$ and $E_i^L$ are the
transverse and longitudinal components of the electric field, and they are 
defined in terms of the projectors $P_{ij}^T$ and $P_{ij}^L$ as 
\be{}
E_i^T=P_{ij}^T E_j \equiv \left(\delta_{ij} - \frac{k_i k_j}{k^2}\right)E_j 
\quad\quad,\quad\quad
E_i^L=P_{ij}^L E_j \equiv \frac{k_i k_j}{k^2}E_j
\ee 
The functions  $\epsilon_T(\omega,k)$ and $\epsilon_L(\omega,k)$ are obtained 
in linear response theory by applying an external electromagnetic field potential $A_j$ and 
measuring the electromagnetic current $J_i$: 
$J_i = q^2 ~G_{ij} A_j$, where the $G_{ij}$ is the retarded current correlator in the medium
and $q$ is the four-dimensional EM coupling.
As we show in the Appendix
\ref{app1}
the transverse and  longitudinal permittivity are 
related to the  transverse and longitudinal
retarded current correlator by
\be{} \label{formulaepsilon}
\epsilon_T(\omega,k) = 1 - \frac{4 \pi }{\omega^2}~ q^2~ G_T(\omega,k)
\quad\quad, \quad \quad
\epsilon_L(\omega,k) = 1 - \frac{4 \pi }{\omega^2} ~q^2 ~G_L(\omega,k)
\ee
Thermodynamical stability requires the positivity of the outgoing heat flow.  This is
associated to the quantities 
\be{}
Q_{\text{heat}} = \frac{\omega}{8\pi} \hbox{Im}({\epsilon_T} (\omega,k)) |E_T|^2
\quad \quad,\quad \quad
Q_{\text{heat}}  = \frac{\omega}{8\pi} \hbox{Im}({\epsilon_L} (\omega,k)) |E_L|^2
\ee
for transverse and longitudinal waves respectively.
They are positive because the function  $\epsilon_T$ and
$\epsilon_L$ are computed from the retarded Green functions 
as in (\ref{formulaepsilon}) and both $~$Im$(G_T(\omega,k))<0$ and Im$(G_L(\omega,k))<0$,
which implies Im$(\epsilon_T(\omega,k))>0$ and  Im$(\epsilon_L(\omega,k))>0$.

Using Maxwell equation (\ref{Maxwell}),
the dispersion relations for the transverse and  longitudinal waves read
\be\label{disprel}
\epsilon_T(\omega,k) = \frac{k^2}{\omega^2}
\quad \quad, \quad \quad
\epsilon_L(\omega,k) = 0
\ee
Due to spatial dispersion $\epsilon_T(\omega,k)$ and $\epsilon_L(\omega,k)$ are functions of both
$\omega$ and $k$.  
Differently from \cite{Amariti:2010jw}, here we do not expand at $k^2$ order. 
In such a way we can capture the effects of 
large spatial dispersion, i.e. we allow for large momentum
dependence in the permittivity.
By using the dispersion relation (\ref{disprel}) one can find 
the on-shell ratio $k^2/\omega^2$ and from this ratio the refractive index $n^2(\omega)$.
This is in general a complex quantity that takes into account the dissipation
of the medium.
Note that, since the permittivity depends on the wave vector, the dispersion
relation can give in general more than one solution for $n^2(\omega)$. 

For a propagating electromagnetic wave,
the sign of the refraction is given by the reciprocal sign of the phase velocity
(Re$[n]$) and 
of the flux of energy (the Poynting vector).
In presence of strong spatial dispersion and strong dissipation, 
the definition of the Poynting vector
and  the determination of the flux of energy
are not well established \cite{Agranovich}.
Hence in this case 
the sign of the refractive index cannot be study 
without the knowledge of a microscopic description of the system.
We discuss about this issue in the Appendix \ref{app2}.
\section{Pekar's theory and ALW}
\label{sec2}

In 1957 Pekar \cite{Pekar} discussed the existence of an
additional light wave in the optical spectra of crystals 
in presence of strong spatial dispersion.
This phenomenon is associated to the presence of a 
resonance in the crystal, namely an exciton.
See \cite{Cocoletzi} for review.
For example in a semiconductor crystal,
an exciton is associated to the transition of an electron,
from the valence to the conduction band.
When such a transition occurs a positive charge carrier, a hole, remains in the valence band.
This hole and the electron interact  throughout a
Coulomb potential. The bound state generated by this  interaction is the exciton.
This exciton couples to light and  
its coupling can be described at macroscopic level through the dielectric
tensor $\epsilon(\omega,k)$. 

The exciton has the same momentum of the absorbed photon, i.e.
the energy necessary to create the exciton depends on the wave-vector $k$.
This translates in the $k$ dependence of the dielectric tensor.
In the simplest case, a single exciton, with effective mass $M$ and
inverse lifetime $\Gamma$, modifies the transverse conductivity as
\be
\epsilon_T(\omega,k) = \epsilon_0 + \frac{\mathcal{A}}{\omega_0 +  \frac{k^2}{2 M}- \omega - i \Gamma}
\ee
where $\omega_0$ is the 
energy required to create an exciton
without momentum.

On shell the dispersion equation  is
$\epsilon_T(\omega,k) = \frac{k^2}{\omega^2} $.
Solving this equation and using the definition $n^2=\frac{k^2}{\omega^2}$
for the refractive index one finds
\be \label{twoind}
n_{1,2}^2 = 
\frac{w^2 \epsilon_0+2 M (w+i \Gamma -\omega_0)\pm\sqrt{8 \mathcal{A} M w^2+\left(w^2 \epsilon_0	
-2 M (w+i \Gamma -\omega_0)\right)^2}}{2 w^2}
\ee
In \cite{Pekar} Pekar concluded that the two solutions of the
dispersion relation in presence of an exciton  
are associated to two light waves, identically polarized.
If there are multi-excitons resonances (other poles depending on the wave-vector
in the dispersion relation) there is an ALW 
associated to each pole.
These waves are distinguished by their different velocity and damping.
This phenomenon is similar to the birefringence
for anisotropic crystals. The main difference is that 
in the case of birefringence the two waves with different refractive index 
have also different polarization.
Moreover birefringence is strongly related to anisotropy, while 
ALWs appear in isotropic systems too.

ALWs can be studied also in longitudinal waves.
In that case, in absence of 
spatial dispersion, there is no light propagating in the crystal. 
Otherwise if spatial dispersion is taken into account the  
propagation is possible, and every pole involving $k$ 
in $\epsilon_L(\omega,k)$ is associated to 
a light wave.

 \subsubsection*{Propagation of ALW}
The propagation of the ALWs in the crystal does not directly follow
from having multiple solutions to the dispersion relation.
Indeed at frequency far from the excitonic resonance the photon does not mix
with the exciton, and a single light wave propagates.
The problem of propagation of the ALW
is related to the choice of the  boundary conditions.
Indeed in the case of strong spatial dispersion the usual Maxwell
boundary conditions are not enough to 
determine the amplitudes of the reflected and transmitted waves.
Pekar faced the problem of propagation of the ALWs by
introducing an additional boundary condition (ABC). 
There are many possible phenomenological choices
of these boundary conditions, and they are related 
to the currents that flow at the crystal surface.
These conditions  define an effective refractive index 
which is used to study the propagation
of the ALWs in the material. This index 
selects which one of the light waves 
propagates at every frequency.

Here we discuss the effective index obtained from the Pekar's
boundary condition.
This condition is $P_{ex}=0$, and it represents 
the continuity of the excitonic polarization at the crystal surface.
If there are two light waves 
the effective refractive index is defined
by the relation
\be \label{Pekind}
n_{\text{eff}} = \frac{n_1}{1-Q}+\frac{n_2}{1-1/Q}
\ee
where $Q=-E_1/E_2$, and $E_1$ and $E_2$ are the complex amplitudes
of the two waves.
From Pekar's ABC it follows that 
$Q=(\epsilon_0-n_1^2)/(\epsilon_0-n_2^2)$
 and
\be \label{pekarABC}
n_{\text{eff}}=\frac{\epsilon_0+ n_1 n_2}{n_1+n_2}
\ee
By changing the ABC 
the behavior of the ratio $Q$ and  of the effective refractive index
changes. Indeed a different choice of 
the ABC leads to a different choice of
$E_1$ and $E_2$, and it affects the propagation.
There are many controversial
issues related to the consistency of the ABCs \cite{Hopfield,Henneberger,Richter,Halevi}. 
In the rest of the paper we assume that 
in our case it is possible to define   $n_{eff}$ 
as in (\ref{pekarABC}).
The effective index $n_{eff}$ determines
which of the two waves is propagating or if
both of them propagate as follows.

If the effective index 
$n_{eff}$ corresponds to one light wave at lower frequencies 
and to the other at higher frequencies, there is 
a frequency regime in which $n_{eff}$  switches from
one index to the other.
This means that both the waves are propagating
in that frequency range.
In this case the KK relations are not 
satisfied separately on the two solutions 
of (\ref{twoind}). The standard optics 
cannot be applied and the correct description is furnished by
Pekar's theory. The effective index, indeed, satisfies the KK relations
and solves the apparent violation of causality.

Standard optics is recovered if the lifetime of the 
exciton is too short.
Indeed in this case
the exciton does not live enough to modify the linear response of the system
and a single wave propagates.
This possibility is theoretically predicted in Pekar's theory because
the lifetime of the exciton resonance varies with the temperature.
If the temperature increases up to a critical value $T_c$, also the inverse lifetime 
increases, and it reaches a critical value $\Gamma_c$.
Below $T_c$ the Pekar's theory must be applied, 
and there is a range of frequencies in which  
both the waves are propagating.
Above $T_c$ the standard optics holds,
and the effective index $n_{eff}$
coincides with only one of the two solutions.

\section{ALW from hydrodynamics}
\label{sec3}

In this section we observe that the propagation of an ALW is possible in media described by 
relativistic hydrodynamics if the transverse current, that couples to the external EM field, has a retarded correlator dominated by a diffusive pole.

Let us consider a system with a $U(1)$ conserved current $J$ which has  a diffusive behavior 
in its transverse part $J_T$: 
$(\partial_t - \mathcal{D} \nabla^2) J_T = 0 $, where $\mathcal{D}$ is the diffusion coefficient. 
The retarded Green function of $J_T$ has the pole  
\be\label{hydro}
G_T({\omega,k}) = \frac{i \mathcal{B} \omega}{i \omega - \mathcal{D} k^2}
\ee
with $\mathcal{B}$  real, up to higher order in $\omega$, $k^2$. 
The analysis of the hydrodynamic equations for a relativistic system at finite 
temperature and chemical potential \cite{Kadanoff,ref2hartnoll} 
shows that the transverse current indeed satisfies a diffusion equation;
the constants $\mathcal{B}$ and $\mathcal{D}$ are related to 
the values of the transport coefficients and thermodynamical quantities by
\be \label{coeffi}
\mathcal{B} =  \frac{\rho^2}{\varepsilon+P}, \quad \quad
\mathcal{D} = \frac{\eta}{\varepsilon+P}
\ee
$\varepsilon$ is the energy density, $\rho$ the charge density, $P$ the pressure and $\eta$
the shear viscosity.
If the retarded correlator of the current is dominated by the pole 
(\ref{hydro}), the transverse permittivity is
\be \label{geneps}
\epsilon_T (\omega,k)= 1-i \frac{4 \pi q^2 \mathcal{B}}{\omega  (i \omega - \mathcal{D} k^2)}
\ee
where $q$ is the EM coupling. By solving the dispersion equation for 
(\ref{geneps}) we find two different
solutions for $n^2$
\be
n_{1,2}^2 = \frac{
\mathcal{D}+i \omega \pm 
\sqrt{
\frac{16 i \pi q^2 \mathcal{D}\mathcal{B} +(\mathcal{D} - i \omega)^2 \omega}{w}
}}
{2 \mathcal{D}}
\ee
There are two different refractive indexes associated to
two different light waves. 
The appearance of multiple solutions to the
dispersion relation may give origin to the propagation of 
ALWs, like in Pekar's theory.
The computation of the transport coefficients is crucial for the study of the existence 
and of the propagation of the ALWs  in a system described by relativistic hydrodynamics.

\section{ALW in holographic optics}
\label{sec4}

As we discussed above the analysis of ALWs in hydrodynamics requires the
knowledge of the  transport coefficients.
There is a class of strongly coupled media
in which they are exactly calculable,
because the Green function 
follows from the application of the holographic principle in string theory.
Indeed the holographic correspondence in string theory implies 
that a strongly coupled quantum field theory  
in $d$-dimensional Minkowski space-time is
equivalent to classical gravity in AdS$_{d+1}$. 
In term of this gauge/gravity duality 
the order parameter of the  $d$-dimensional field theory are
related to the order parameter of a black hole solution in AdS$_{d+1}$.
In this paper we focus on the case $d=4$.
In the hydrodynamical regime
we can analitically
compute the EM response functions of this 
 strongly coupled 
system,
 at finite temperature and charge density.
If these response functions have one or more 
poles in the $(k,\omega)$ plane we can
observe the generation of one or more ALWs.

In crystal optics the ALWs are the consequence of 
the strong spatial dispersion, which 
is associated to the excitonic resonances. The interaction
of the excitons  
with the photon is described at macroscopic level by 
the presence of particular poles in the electromagnetic 
response function $\epsilon(\omega,k)$.
In the holographic correspondence we know that the 
poles of the response functions are associated to
quasi normal frequencies in AdS gravity \cite{Kovtun:2005ev}.

Quasi normal modes are solutions 
to the linearized
equations of the classical fluctuations 
of the black hole with ingoing boundary conditions.
Ingoing boundary conditions are chosen 
because classically  the horizon does not emit radiation.
The classical fluctuations 
are interpreted 
as small deviations 
from thermodynamic equilibrium
in the dual field theory,
and the incoming boundary conditions
correspond 
to dissipation.
The dispersion relations of this dissipative process
are encoded in the poles of the retarded Green function,
which are located at the (complex) eigenfrequencies of the
gravitational quasi normal modes.
 
We conclude that the quasi normal frequencies in 
AdS gravity are the holographic 
duals of the
excitonic resonances in the strongly
coupled field theory.

In this section we study the propagation of light waves both for transverse and 
for longitudinal waves in media described by a dual gravitational background, in
the hydrodynamical regime
\footnote{Observe that 
the calculation is valid if  $|k| =|n| \omega$ is still
hydrodynamical}.
In the simplest case, at zero charge density but at finite temperature, the transverse
propagator does not posses any pole, 
and there is only one  propagating light wave in the medium.
At finite charge density a diffusive pole is generated, and indeed this corresponds
to a new branch for the refractive index $n^2$, which is associated to the propagation of an
ALW.

In the longitudinal case, 
at zero charge density but finite temperature,  the correlator  
has a single diffusive pole, and there is
a single light wave.
Giving a non zero value to the chemical potential produces another  pole: the second sound.
The emergence of this new quasi-normal mode
generates the second solution of $n^2$ which is associated
to an ALW.

\subsection{Transverse waves without chemical potential: single wave}
\begin{figure}[ht]
\begin{minipage}[b]{0.5\linewidth}
\includegraphics[width=7cm]{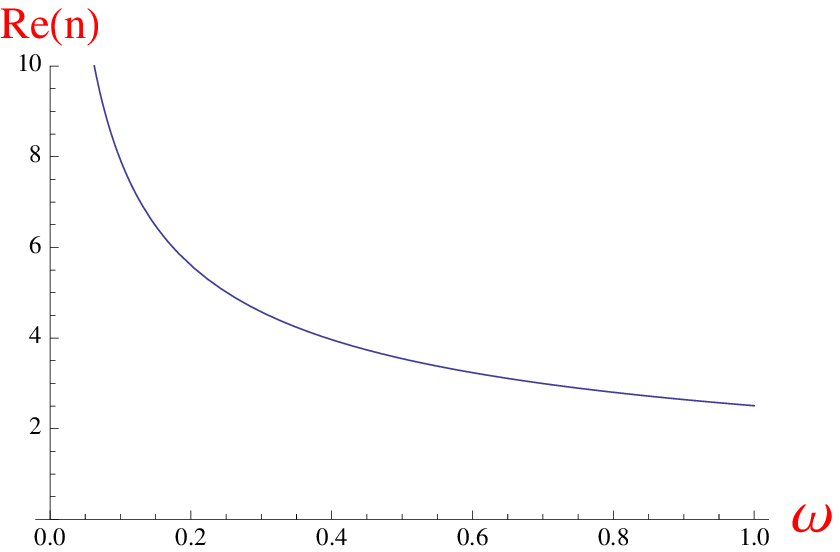}
\caption{\emph{Re$(n)$ for transverse waves 
$~~~~~$ with
$~T=1$, $\mu=0$  and $q=0.1$.}}
\label{fig1}
\end{minipage}
$~~~~$
\begin{minipage}[b]{0.5\linewidth}
\includegraphics[width=7cm]{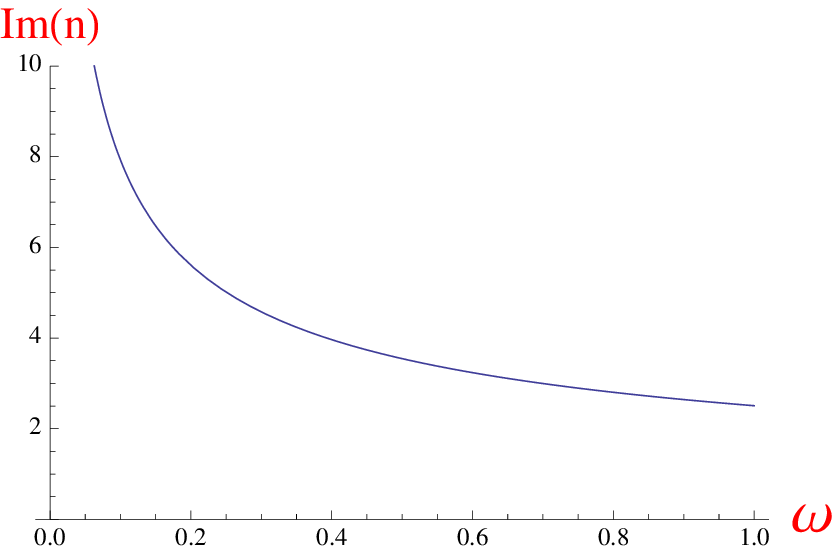}
\caption{\emph{Im$(n)$ for transverse waves$~~~~~~$ with
$~T=1$, $\mu=0$  and $q=0.1$.}}
\label{fig2}
\end{minipage}
\end{figure}
We concentrate on the properties of a medium described by its 3+1 dimensional 
energy momentum tensor $T^{\mu\nu}$ and its 
3+1 dimensional $U(1)$ conserved current $J^{\mu}$.
This subsector of the 3+1 dimensional medium is encoded in a 5 dimensional 
metric $g_{mn}$ and a five dimensional gauge field $A_m$, with the action:
\be{}\label{azione}
S = \frac{1}{2 e^2 l^2} \int d^5 x \sqrt{-g} \left(R- \frac{6}{l^2} \right)-\frac{1}{4 e^2} \int d^5 x \sqrt{-g} F_{mn}F^{mn}
\ee
where $e$ is the five dimensional EM coupling constant and $3/l^2$ is the cosmological constant. We consider the 3+1 dimensional plasma in an homogeneous and isotropic state that is in thermal equilibrium, with zero charge density. This ground state of the system is described by a particular solution of (\ref{azione}), an uncharged AdS black hole: 
\be{} 
d s^2 = \frac{l^2}{4 b^2u}\left( d x^2+ d y^2+dz^2-f(u)dt^2\right)
+\frac{l^2}{4 u^2 f(u)}du^2
\ee
where $b=\frac{1}{2 \pi T}$ and  $f(u) = 1-u^2$. The black hole horizon is at 
$u=1$ and the AdS boundary is at $u=0$. To compute the linear response of the 
system at an external electric perturbation we need to linearize the 
equation of motion for the component of the gauge field $A_m(u)e^{-i\omega t +i k z}$ parallel to the 
wave vector $k$. We need then to solve these equations with the infalling 
boundary condition at $u=1$ and arbitrary Dirichlet boundary conditions at $u=0$.

The retarded correlator for the  transverse current has been computed  
in \cite{Policastro:2002ne} at small frequencies and wave vector:
\be{}\label{corr}
G_T(\omega,k) = -\frac{  i \omega }{4 b}
\ee
with $b=\frac{1}{2 \pi T}$. In this case transverse permittivity $\epsilon_T(\omega,k)$ is
\be{}
\label{epstrasvzero}
\epsilon_T (\omega ,k)=\epsilon_T (\omega )=1+\frac{ i \pi q^2}{b \omega}
\ee
The holographic calculation leaves an ambiguity in the definition of the correlator: a term proportional to $(\omega^2 - k^2)$. The value of this term has to be fixed by the 
holographic renormalization procedure \cite{Skenderis}. The physical requirement 
we use is that at large $\omega$ the permittivity goes to its vacuum value:
$\epsilon(\omega \rightarrow \infty)=\epsilon_T(\omega\rightarrow \infty,0) \rightarrow 1$. 
We computed numerically the large $\omega$ behavior of the correlators and we fixed 
in this way the value of the renormalization constant.
Anyway, we observe that all our results are qualitatively independent of this renormalization constant.

Spatial dispersion  in (\ref{epstrasvzero}) does not give 
any effects, and a single wave
is expected. Indeed the squared refractive index is given by
\be
n^2 = 1+\frac{i \pi  q^2 }{b \omega }
\ee
and the refractive index, which real and imaginary part are plotted in Figures \ref{fig1} and \ref{fig2},
is
\be
n= \sqrt{1+\frac{\pi^2 q^4}{ 2 \, b^2 \omega^2}}+ i \frac{\pi q^2}{\sqrt 2 b \omega}
\ee
As explained in the Appendix \ref{app2},
in this case  the sign of the refractive index can be studied by comparing Re[$n(\omega)$] with the Poynting vector.
As claimed in \cite{Amariti:2010jw} the light wave is positively refracted. 
\subsection{Transverse waves with chemical potential: ALW}
\begin{figure}[ht]
\begin{minipage}[b]{0.5\linewidth}
\includegraphics[width=7cm]{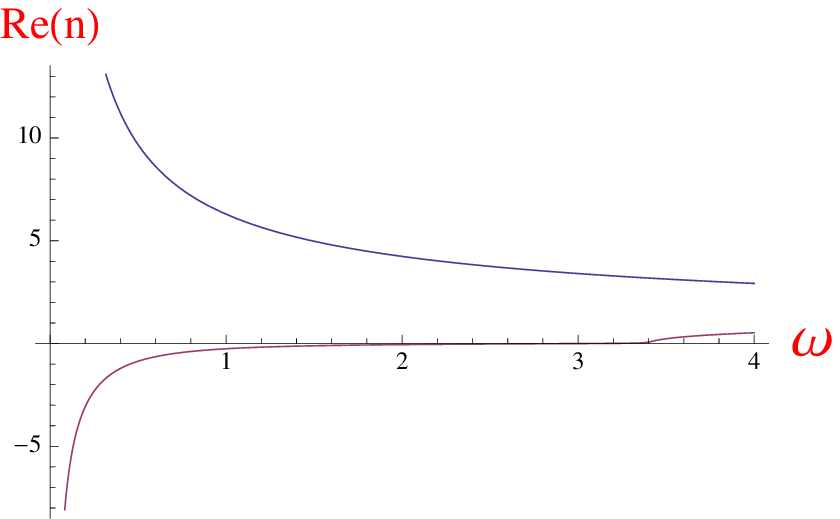}
\caption{\emph{Re$(n)$ for transverse waves $~~~~~$
with  $\mu/T=10$ and $q=0.1$.}}
\label{fig3}
\end{minipage}
$~~~~$
\begin{minipage}[b]{0.5\linewidth}
\includegraphics[width=7cm]{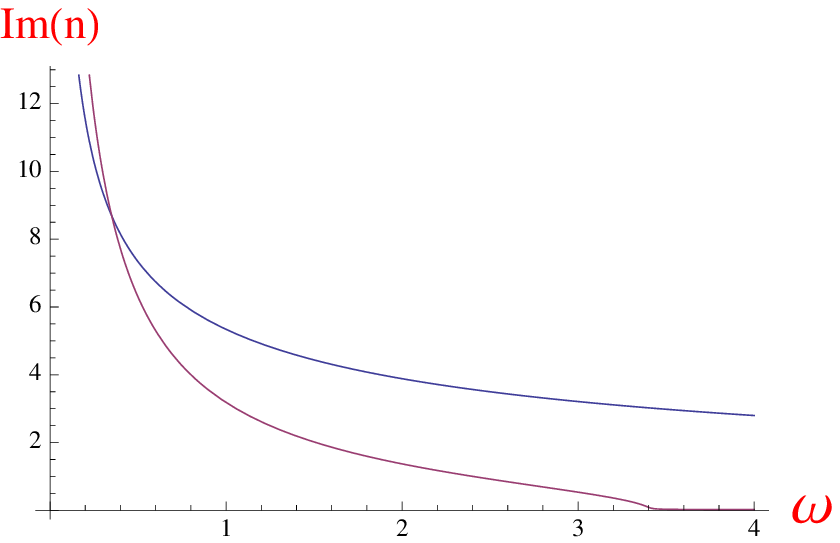}
\caption{\emph
{Im$(n)$  for  transverse waves $~~~~~$ with $\mu/T=10$ and $q=0.1$}}
\label{fig4}
\end{minipage}
\end{figure}
In this section we study a
3+1 dimensional medium at finite charge density. 
This ground state of the system is described by a different solution of (\ref{azione}), 
a charged AdS-RN black hole, with metric and gauge field given by
\begin{eqnarray}\label{RN}
&& ds^2= \frac{(2-a)^2 l}{16 \, b^2}\frac{1}{u} \left(\text{d}x^2+\text{d}y^2+\text{d}z^2-f(u) \text{d}t^2 \right)+\frac{l^2}{4 } \frac{\text{d}u^2}{u^2 f(u)} \nonumber \\
&& A_t= - \frac{u}{2 b} \, \sqrt{\frac{3}{2}a}+\mu
\end{eqnarray}
where
\be
f(u) = (1-u)(1+u - a u^2)
\ee 
As before $u=0$ and $u=1$ correspond respectively to the AdS boundary
and to the black hole horizon.  
The temperature $T$ and the chemical potential $\mu$ are
given in terms of the  parameters $a$ and $b$ as
\be
T = \frac{2-a}{4 \pi b} \quad , \quad \mu = \sqrt{\frac{3}{2} a} \frac{1}{2 b}
\ee
The retarded correlator for the  transverse current has been computed  
\cite{Ge:2008ak}  at small frequencies and wave vector:
\be{}
G_{T}(\omega,k)=
\frac{3 a}{4 (1+a) b^2}\left(\frac{i\omega}{-D k^2+i \omega}\right)-\frac{(2-a) i \omega}{8 (1+a)^2 b }
\ee
where $D = \frac{2-a}{4 \pi T}$

This case corresponds to the one studied in \cite{Amariti:2010jw}.  
In that case we expanded for small $k$ and we ignored 
strong effects coming from spatial dispersion.
In that approximation the region of very small frequency 
cannot be explored, and we found only one wave
(with negative refraction) propagating in the system. 
In order to study the effects of strong spatial dispersion, we should
not expand in the momentum. The transverse permittivity is then given by
\be
\epsilon_T(\omega,k)=1+\frac{4 i \pi q^2}{\omega}
\left( \frac{3 a}{4 (1+a) b^2}\left(\frac{1}{D k^2-i \omega}\right)+\frac{(2-a) }{8 (1+a)^2 b }\right)
\ee
whose pole structure is analogous to the one studied in section \ref{sec3}.
Indeed, by solving the dispersion relation $\epsilon_T=n^2$,
we obtain two different solutions for $n^2$.
One of the solution corresponds to the one studied in \cite{Amariti:2010jw}. The other 
instead is the new solution (the ALW) associated to the strong spatial dispersion effects.
We plot in blue and in red the refractive index of the two waves: in Figure \ref{fig3} the 
real part while in Figure \ref{fig4} its imaginary part.

To conclude this section we comment on the sign of the refractive index for the light
waves. Both light waves are strongly dissipative, as can be seen in
Figure \ref{fig4}. 
The wave corresponding to the red line in the Figure \ref{fig3} is the one studied
in \cite{Amariti:2010jw}. There we studied a frequency window where
 the spatial dispersion effects are small. The Poynting vector can
 then be easily determined and we were able to prove that there is negative refraction.
 The other wave (blu line in Figure \ref{fig3}) is everywhere in frequencies characterized
 by strong spatial dispersion. Hence the flux energy direction cannot be easily determined
 and as a consequence we need additional informations on the microscopic model
 to study the sign of the refraction (see appendix \ref{app2}). 
\subsection{Longitudinal waves without chemical potential: single wave}

\begin{figure}[ht]
\begin{minipage}[b]{0.5\linewidth}
\includegraphics[width=7cm]{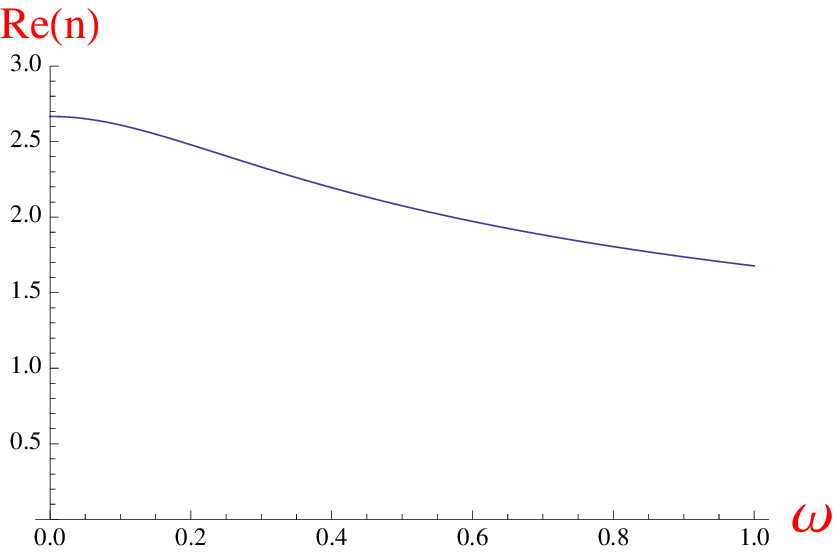}
\caption{\emph{Re$(n)$ for longitudinal waves
$~~~$ with $T=1$, $\mu=0$  and q=0.1.}}
\label{fig5}
\end{minipage}
$~~~~$
\begin{minipage}[b]{0.5\linewidth}
\includegraphics[width=7cm]{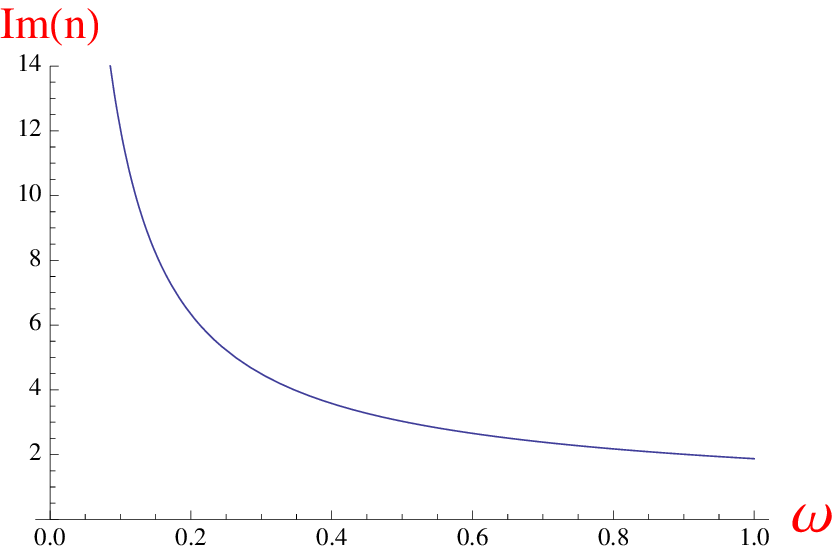}
\caption{\emph{Im$(n)$ for longitudinal waves
$~~~$  with $T=1$, $\mu=0$  and q=0.1.}}
\label{fig6}
\end{minipage}
\end{figure}

At finite temperature light waves propagate also in the longitudinal channel.
Usually longitudinal waves do not propagate in the vacuum because the
dispersion relation,  $ \epsilon_L(\omega)=0$, does not generate any refractive index.
Nevertheless at finite temperature the background 
studied above gives origin to a diffusive pole
in the current correlator, even in absence of 
finite charge density.  This pole in the $(k, \omega)$ 
plane implies that a solution to the dispersion relation exists
and there is a propagating light wave in the longitudinal channel.

The retarded correlator for the longitudinal current has been computed  
in \cite{Policastro:2002ne} at small frequencies and wave vector:
\be{}\label{corrL}
G_L(\omega,k) = \frac{l}{2  e^2 b} \quad 
\frac{\omega^2} 
{ i \omega - k^2 b }
\ee
In this case the longitudinal permittivity $\epsilon_L(\omega,k)$ is
\be{}
\epsilon_L (\omega,k )=1- \frac{2 \pi q^2 l }{e^2 b(i \omega - k^2 b)}
\ee
The holographic calculation leaves an ambiguity in the definition of the correlator: a term proportional to $\omega^2$. The value of this term has to be fixed by the 
holographic renormalization procedure \cite{Skenderis}. We use the 
same requirement of the previous section: $\epsilon(\omega\rightarrow\infty)=\epsilon_L(\omega\rightarrow\infty,0) \rightarrow 1$. The large $\omega$ 
behavior of the correlators has been computed numerically, and this fixes the value of the 
renormalization constant.
Once again we observe that our results do not depends qualitatively on this renormalization 
constant.
The refractive index 
is computed by solving the dispersion relation 
$\epsilon_L(\omega,k)=0$.
There is only a single wave propagating in the medium.
In Figure \ref{fig5} and Figure \ref{fig6} we plotted the real and 
imaginary part of the refractive index. 
\subsection{Longitudinal waves with chemical potential: ALW}
\begin{figure}[ht]
\begin{minipage}[b]{0.5\linewidth}
\centering
\includegraphics[width=7cm]{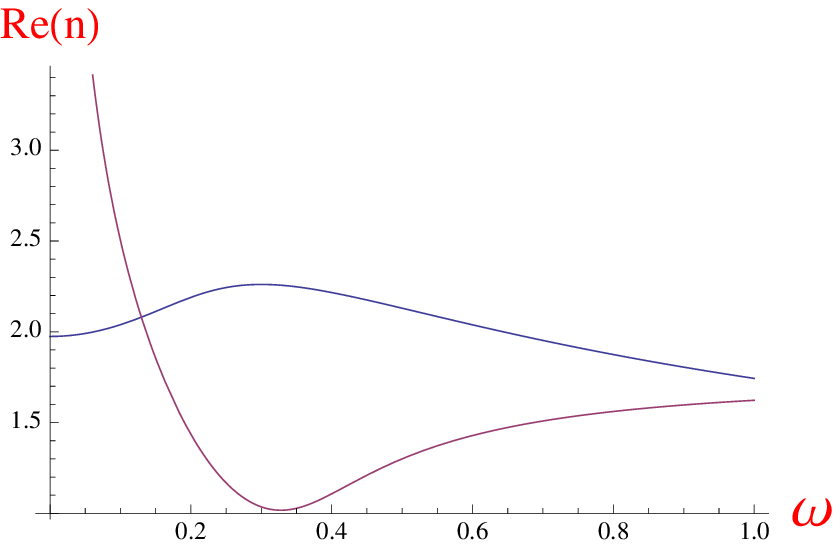}
\end{minipage}
\hspace{0.5cm}
\begin{minipage}[b]{0.5\linewidth}
\centering
\includegraphics[width=7cm]{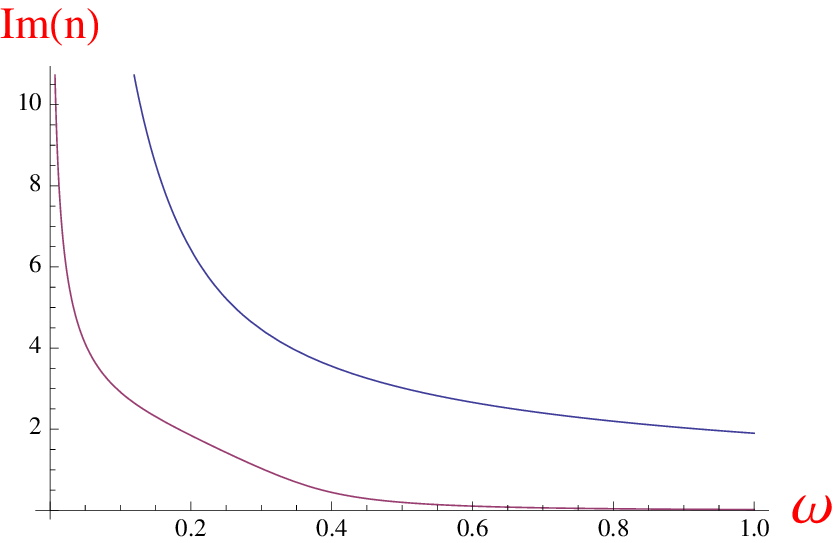}
\end{minipage}
\\
\begin{minipage}[b]{0.5\linewidth}
\centering
\includegraphics[width=7cm]{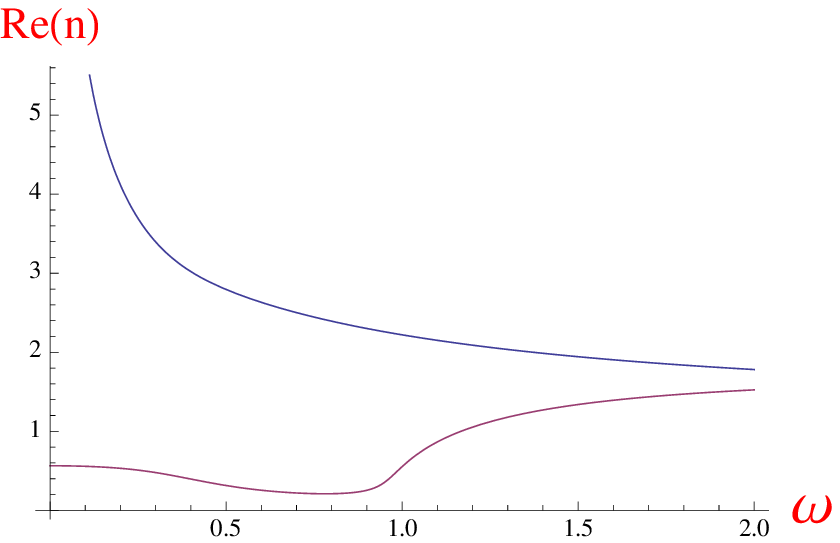}
\end{minipage}
\hspace{0.5cm}
\begin{minipage}[b]{0.5\linewidth}
\centering
\includegraphics[width=7cm]{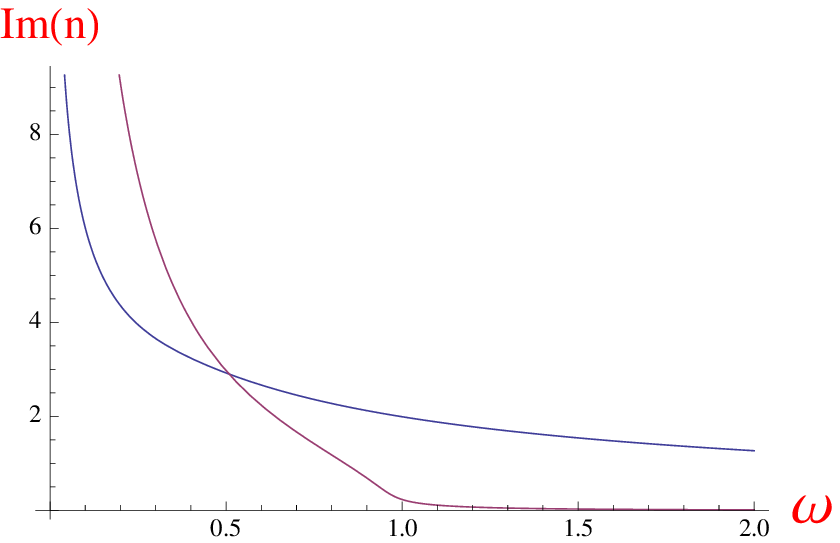}
\end{minipage}
\caption{\emph
{Re$(n)$ and Im$(n)$  for longitudinal waves,  with  
$\mu/T=1$ (above) and $\mu/T=3$ (below) and $q=0.1$.}}
\label{figlong}
\end{figure}
Here we proceed by analyzing the optics of the
 system when we add a chemical potential.
At finite charge density there is another pole, 
corresponding to a new collective mode.
This mode, called the second sound, is associated
to another excitonic resonance.
Both resonances are peaked around the same frequency $\omega \simeq 0$,
and in this case there are two different light waves, two 
mixing states of the photon with the excitons, propagating in the medium.

The retarded correlator for the longitudinal current has been computed  
\cite{Matsuo:2009yu} at small frequencies and wave vector:
\be{}\label{Gzzchemical}
G_{zz}(\omega,k)=
-\frac{w^2 }{4 (1+a) b^2}\left(\frac{9 a}{k^2-3 \omega ^2}+\frac{2 (2-a)^2 b}{2 (2+a) b k^2-4 i (1+a) \omega -(2-a)^2 b D_m \omega ^2}\right)
\ee
with
\footnotesize
\be{}
D_m=\frac{2 \left(4 (1+a)^3 (1+4 a)^2 \log (2\!-\!a)\!-\!27 (2-a) a^2 (1+4 a)+2 (1+a)^2 \sqrt{1+4 a} 
(2+a (41 a\! -\!2)) \log \left(\frac{3+\sqrt{1+4 a}}{3-\sqrt{1+4 a}}\right)\right)}{(2-a)^4 (1+4 a)^2}
\ee
\normalsize
The sound pole $(k^2-3 w^2)$ leads to a second 
solution for the dispersion relation and hence to a second
light wave.
We fix $T=1$ and vary the ratio $\mu/T$, from low to larger values.
We plot the real and the imaginary part of the refractive index in Figure \ref{figlong},
for $\mu/T=1$ and for $\mu/T=3$.
From the figure we observe that the two solutions $n_1$ and $n_2$ 
intersect at small frequency for small chemical potential. 
They do not intersect anymore
for larger values of $\mu$.
This effects is associated to the transition between 
optical regimes, where the  number of propagating
waves changes.
We give a more detailed analysis in next section.

\section{Propagation of the ALW in our media}
\label{sec5}

In this section we discuss the propagation of the ALWs in our media.
At finite temperature and charge density 
we have found more solutions to the dispersion relation,
leading to ALWs.
The propagation of ALWs is related to the boundary condition,
as explained in section \ref{sec2}.

An important difference between the situation that
we studied here and the case of crystals, for
which the Pekar's' theory has been developed,
is that, here, we deal with an infinite medium.
In crystal optics the ABCs are imposed at the surface of the 
medium. Here we assume that similar conditions can be imposed 
at infinity such that it is possible to define  $n_{eff}$ 
as in (\ref{pekarABC}), with $\epsilon_0=1$. 

\begin{figure}[ht]
\begin{minipage}[b]{0.5\linewidth}
\centering
\includegraphics[width=7cm]{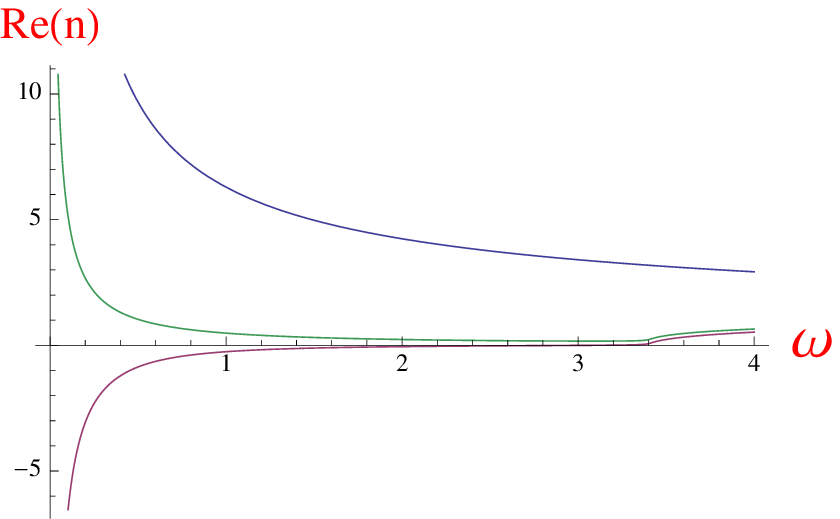}
\caption{\emph{$n_{eff}$ for the transverse waves 
  $~~~~~~$
  at
$\mu/T=10$ and $q=0.1$.}}
\label{figtraa}
\end{minipage}
\begin{minipage}[b]{0.5\linewidth}
\centering
\includegraphics[width=7cm]{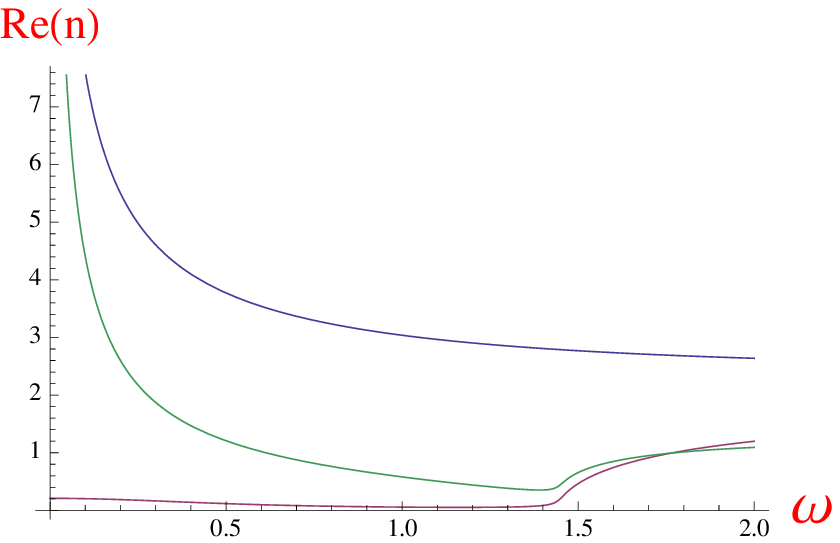}
\caption{\emph
{$n_{eff}$ for the longitudinal waves
at $\mu/T=5$ and $q=0.1$.}}
\label{figlona}
\end{minipage}
\end{figure}

We plot the effective index $n_{eff}$ 
for  the transverse waves in Figure  \ref{figtraa}.
From the figure we observe that $n_{eff}$ switches from $n_1$ to $n_2$ 
at small frequencies, near the resonance. 
By borrowing the crystal optics physical interpretation, 
the behavior of $n_{eff}$ suggests that 
both the light waves propagate.
\begin{figure}[ht]
\begin{minipage}[b]{0.5\linewidth}
\centering
\includegraphics[width=7cm]{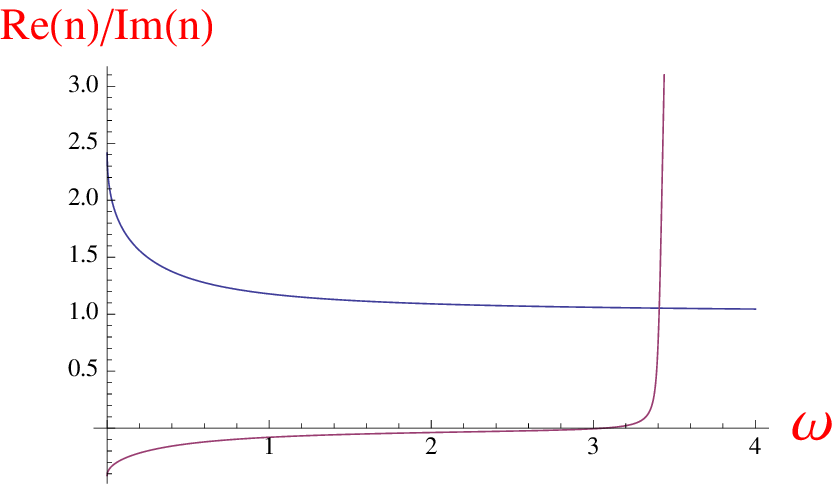}
\caption{\emph{Ratio $Re(n)/Im(n)$ for the $~~~$
transverse waves at $\mu/T=10$ and $q=0.1$.}}
\label{figtr}
\end{minipage}
\begin{minipage}[b]{0.5\linewidth}
\centering
\includegraphics[width=7cm]{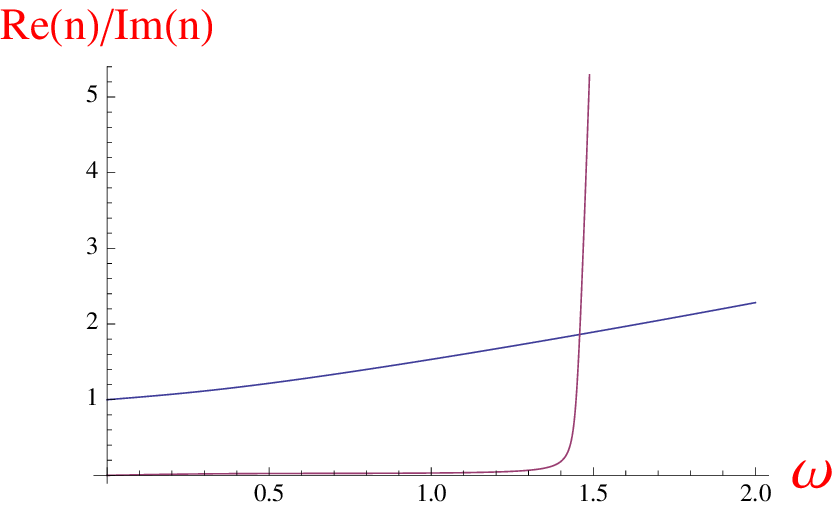}
\caption{\emph
{Ratio $Re(n)/Im(n)$  for the longitudinal waves at $\mu/T=5$ and $q=0.1$.}}
\label{figlon}
\end{minipage}
\end{figure}
We can imagine an alternative check of this result by looking at the propagation of
each light wave with respect to the dissipation. This property is encoded in the 
ratio $Re(n)/Im(n)$.
We plotted this ratio in Figure \ref{figtr}.
At low frequency both the waves are strongly dissipative, and the
ratio $Re(n)/Im(n)$ is comparable for them. As the frequency increases 
the wave selected by $n_{eff}$ is preferred because 
the absorption is lower for that one. 

An analogous phenomenon is found 
in the longitudinal 
waves. We plot in Figure 
\ref{figlona} the effective index and in Figure 
\ref{figlon} the ratio $Re(n)/Im(n)$ .

We conclude that both in the transverse and in the longitudinal channel we have
propagating ALWs.
\subsection{Transitions between different optical regimes}

As explained in section \ref{sec2}, in 
Pekar's theory for crystals, 
a transition from propagating ALWs
to standard optics  takes places at some 
critical value of the order parameter,
where the lifetime of the exciton become short.

Here, in the transverse case, at finite temperature and charge density,
this transition does not occur.
At every value of the order parameters
there is a range of frequency where 
both waves propagate.
Indeed $n_{eff}$  always switches from $n_1$
to $n_2$ for every  temperature and chemical potential.

\begin{figure}[ht]
\begin{minipage}[b]{0.5\linewidth}
\centering
\includegraphics[width=7cm]{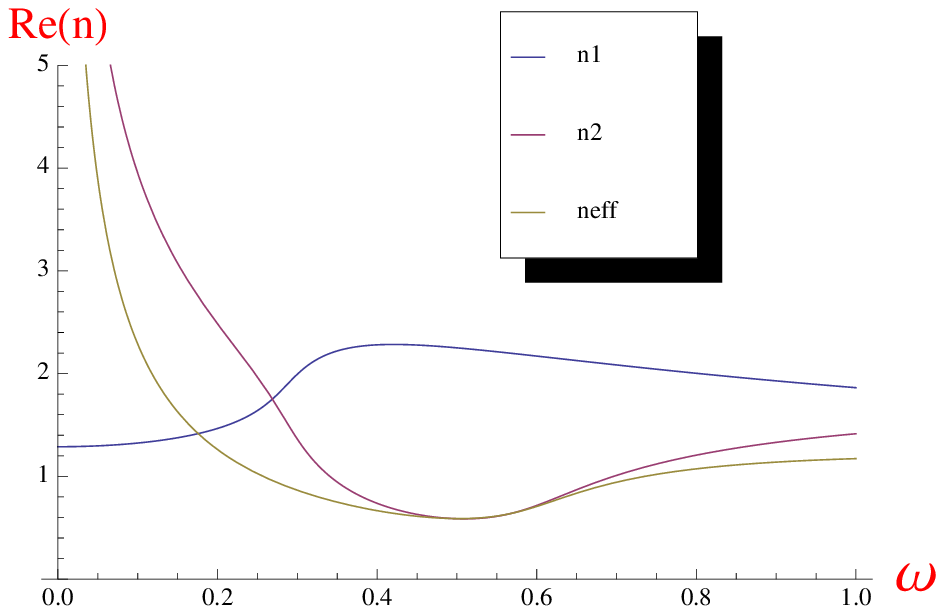}
\end{minipage}
\hspace{0.5cm}
\begin{minipage}[b]{0.5\linewidth}
\centering
\includegraphics[width=7cm]{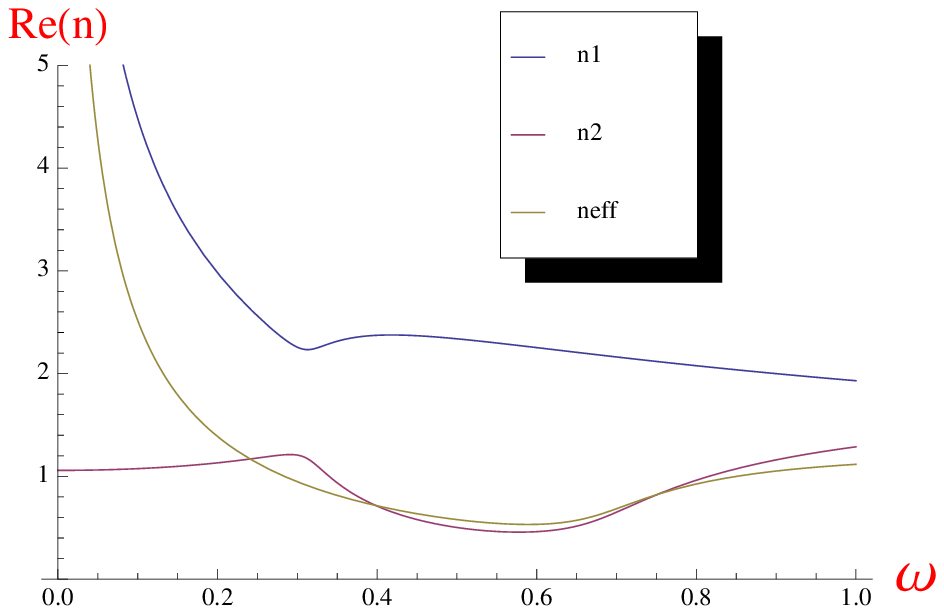}
\end{minipage}
\caption{
\emph{Optical transition for $T=1$ and $q=0.1$, 
at $\mu=\mu_{crit}$ for longitudinal waves: for $\mu<\mu_c$ only one light wave
is propagating, while for $\mu>\mu_c$ both waves propagate at low frequency.}}
\label{figtrans}
\end{figure}

In the longitudinal case the 
transition to standard optics 
cannot exist,
because,
without taking into account spatial dispersion, 
there are no light waves
propagating in the longitudinal channel.
Nevertheless we observe here a transition between different optical regimes,
as can bee seen in Figures \ref{figtrans}.
For low values of the chemical potential,
$n_{eff}$ coincides
with a single light wave in the whole hydrodynamical regime.
In this case this is the only light wave propagating in the medium.
If the chemical potential increases it reaches a critical value $\mu_{crit}$.
For $\mu>\mu_{crit}$ the effective index $n_{eff}$ 
switches from $n_1$ to $n_2$.
 In the frequency range where the transition of $n_{eff}$ takes place
there are two propagating light waves.

The physical 
explanation of the transition is that
the chemical potential is an order parameter for the mixing of the two 
excitons with the photon.
For small values of $\mu$ there is only one light propagating in the medium,
and there is no large mixing between the two 
excitons.
The single propagating wave is associated to the sound pole in  (\ref{Gzzchemical}),
 which is not dissipative and it is the dominant one at low $\mu$.
For larger $\mu$ there is larger mixing between the two quasi-normal modes,
and both can propagate, depending on the frequency.
\section*{Conclusions}
\addcontentsline{toc}{section}{Conclusions}
\label{conc}

In this  paper we  observed that 
theories described by relativistic hydrodynamics
with a diffusive behavior for the currents
can give origin to the phenomenon of additional light waves.
This property has been theoretically predicted by 
Pekar in crystal optics and then experimentally verified.
As in the case of crystals, here the ALWs are connected with 
the strong effects of spatial dispersion, i.e. the $k$ dependence of
the poles of the Green functions. 

We verified our conjecture on the propagation of  ALWs 
in media described by relativistic hydrodynamics 
by studying a strongly coupled field theory 
that can be holographically described. In this case the 
Green functions and the transport coefficients are exactly calculable
thanks to the gauge/gravity duality.

Note that 
with our analysis we can only capture the hydrodynamical
regime of the theory. Indeed at larger frequency and momentum there are
other quasi-normal modes. 
They were first computed in \cite{Horowitz:1999jd}
 for black holes in AdS.
These modes exist  only for some discrete complex frequencies 
associated to new poles of the Green functions 
\cite{Birmingham:2001pj,Starinets:2002br,Nunez:2003eq}.
For the  transverse and longitudinal
R-current correlators the quasi-normal modes
have been studied in \cite{Kovtun:2005ev,Teaney:2006nc,Kovtun:2006pf}
 \footnote{See also \cite{Amado:2007yr} for the computation of the residues.}.
In the dual field theory these poles 
correspond to the presence of excitons that
can lead to other ALWs, propagating at the various
resonance frequencies.  
It would be interesting to perform a detailed study of the existence and the
propagation of the ALWs out of the hydrodynamical regime. For this computation the use
of the gauge/gravity correspondence will be fundamental. 

Another issue is the study of the transition to standard optics. In crystal optics
it is driven by the $i \Gamma$ term appearing in the pole of
the dielectric tensor. In our case this term is set to zero in an ideal system, but it
arises if impurities are taken into account. There is a general strategy to compute the
contribution of the impurities to the Green functions in
the gauge/gravity duality \cite{Hartnoll:2008hs}. Once this contribution is added we expect that 
there is some critical value of the order parameters at which the exciton lifetime is
too short for the ALWs to propagate. It would be nice to check this prediction in
explicit models.

This paper is a second step in the understanding of optical properties 
of strongly coupled media via holography. 
Indeed as
in \cite{Amariti:2010jw} 
the gauge/gravity duality is an useful
laboratory to verify some general statements 
common to many systems\footnote{See also \cite{Gao:2010ie}, in which
the authors studied the optical  
properties of an holographic superconductor.}. 
With this motivation many other optical properties of
strongly coupled systems can be  studied.

\section*{Acknowledgments}

It is a great pleasure to thank David Mateos 
for comments on the draft. We also thank  V. M. Agranovich, 
Y. Gartstein, V. Ginis and A. Paredes  for nice discussions.
A.A. is supported by UCSD grant DOE-FG03-97ER40546;
A. M. is a Postdoctoral researcher of FWO-Vlaanderen. A. M. is also supported in part
by the Belgian Federal Science Policy Office through the Interuniversity Attraction Pole
IAP VI/11 and by FWO-Vlaanderen through project 
 G.0114.10N; D. F. is
supported by CNRS and ENS Paris. 

\appendix

\section{$\epsilon_T$ and $\epsilon_L$}\label{app1}

In this Appendix we derive the expression for $\epsilon_T$ and $\epsilon_L$
in term of the transverse and longitudinal correlators.
The relevant equation is
\be{} \label{maxwe}
\partial_t E + 4 \pi J = \partial_t D
\ee
where $D_i = \epsilon_{ij} (\omega, k) E_j$.

The vector potential is $A_i = P_{ij}^T A_i+ P_{ij}^L A_i \equiv A_i^T + A_i^L$.
The field $E_i$ is expressed in term of the vector potential
$A$ as $E_i =F_{0i} = - i \omega A_i- i k A_0$.
From the last equality in (\ref{maxwe}) we have
\begin{eqnarray} \label{base}
4 \pi J_i &=&~~~~\partial_t D_i- \partial_t E_i ~~= - i \omega (D_i-E_i) \nonumber \\&=&
- i \omega (\epsilon_{ij}-\delta_{ij}) E_j =
- \omega (\epsilon_{ij}-\delta_{ij}) ( k_j A_0+  \omega A_j)
\end{eqnarray}
where $k_i=(0,0,k)$.
In the linear response theory the Green function is related to the current through the relation
\be\label{JJ}
J_i = q^2\left( G_{ij} (\omega,k)  A^j + G_{i0}(\omega,k )A^0\right)=q^2\left(G_{ij} (\omega,k)  A_j - G_{i0}(\omega,k )A_0\right)
\ee
We now make use of the identities $G_{z0}=-\frac{k}{\omega} G_{zz}$ and $G_{x0}=G_{y0}=
G_{zx}=G_{zy}=0$.
From  these relations and from  (\ref{base})  and  (\ref{JJ})
we obtain
\be{}
\epsilon_{ij} (\omega,k)= \delta_{ij}- \frac{4 \pi}  {\omega^2} ~q^2~G_{ij}(\omega,k)
\ee
By projecting this relation on the  transverse and longitudinal channels we obtain
\be{}
\epsilon_T(\omega,k) =1 -\frac{4 \pi}{\omega^2} ~q^2~ G_T(\omega,k)\quad \quad, \quad \quad 
\epsilon_L (\omega,k)= 1 -\frac{4 \pi }{\omega^2} ~q^2~ G_L(\omega,k)
\ee

\section{Poynting vector}
\label{app2}

In the paper we showed different physical systems exhibiting the phenomenon
of ALW. It is interesting 
to determine if a propagating wave shows
negative refraction too. 
There is negative refraction when the phase vector is opposed to the energy flux vector.
The sign of the phase velocity is given by Re$\[n(\omega)\]$
while the energy flux is given by the Poynting vector.
However, the Poynting vector is not always well defined \cite{Agranovichlibro}. 

In the case of small dissipation, even if spatial dispersion is present, the
Poynting vector is oriented as the group velocity $v_g$, 
and has the simple expression 
\be
\label{Poyntingspatial}
S = \hbox{Re}\left({\bf E}^* \wedge {\bf B}-\frac{\omega}{2}\frac{\partial \epsilon_{ij}}{\partial {\bf k}} {\bf E}_i^* {\bf E}_j\right)
\ee
Negative refraction in this context has been studied in \cite{Agranovich}.

When dissipation effects are not negligible, the Poynting vector
is not in general directed as the group velocity. 
If spatial dispersion effects are small, we can compute it using the 
classical $\epsilon$, $\mu$ approach
\footnote{Where $\mu$ here refers to the magnetic permeability.}
(see  \cite{McCall,Depine}). 
\be
S = \hbox{Re}\left(\frac{n}{\mu} \right) |E|^2
\ee
In \cite{Amariti:2010jw} we showed that this expression is equivalent to (\ref{Poyntingspatial})
for transverse waves, by expanding the permittivity at second order in the wave vector.
In that case we can studied negative refraction by comparing the sign of
the Poynting vector with the sign of Re$\[n(\omega)\]$.

In the general case of strong spatial dispersion and strong dissipation we cannot
identify precisely the direction of the flux of energy \cite{Agranovichlibro}. Hence we cannot
determine if there is negative refraction.
Except for
the wave also analyzed in \cite{Amariti:2010jw}, all the other waves
we studied in this paper fall in this category.

\end{document}